# Industry Best Practices in Robotics Software Engineering

- **Robert Bocchino** (Jet Propulsion Laboratory, California Institute of Technology, USA)[1]
- **Arne Nordmann** (Robert Bosch Corporate Research, Germany)
- **Allison Thackston** (Waymo, USA)
- **Andreas Angerer** (XITASO GmbH, Germany)
- **Federico Ciccozzi** (Mälardalen University, Sweden)
- **Ivano Malavolta** (Vrije Universiteit Amsterdam, The Netherlands)
- **Andreas Wortmann** (University of Stuttgart, Germany)

**ABSTRACT**

Robotics software is pushing the limits of software engineering practice. The 3rd International Workshop on Robotics Software Engineering held a panel on "the best practices for robotic software engineering". This article shares the key takeaways that emerged from the discussion among the panelists and the workshop, ranging from architecting practices at the NASA/Caltech Jet Propulsion Laboratory, model-driven development at Bosch, development and testing of autonomous driving systems at Waymo, and testing of robotics software at XITASO. Researchers and practitioners can build on the contents of this paper to gain a fresh perspective on their activities and focus on the most pressing practices and challenges in developing robotics software today.

## 1. Introduction

Robots are increasingly entering our lives: cleaning robots are in our homes, (partly) automated vehicles bring us to work where we cooperate with service robots, industrial robots, or agricultural robots. Successfully engineering robots demands the interdisciplinary collaboration of experts from, e.g., electrical engineering, mechanical engineering, artificial intelligence, and software engineering. Consequently, integrating modules contributed by respective experts is a key challenge in engineering software-centric robots, yet it is only one of the cross-cutting software concerns crucial to robotics. Hence, most of the added value contributed by domain experts, as well as the glue between the individual experts' contributions in modern robots, is software.

Successfully engineering, deploying, and operating robotics applications demands suitable software methods for domain experts providing modules, integrating, validating, verifying, deploying, and evolving these. At the 3rd International Workshop on Robotics Software

[1] © 2022 California Institute of Technology. Government sponsorship acknowledged. This research was carried out at the Jet Propulsion Laboratory, California Institute of Technology, under a contract with the National Aeronautics and Space Administration (80NM0018D0004).

Engineering[2] (RoSE), held in June 2021 and co-located with the International Conference on Software Engineering, researchers and practitioners from across the world discussed the challenges, advances, and opportunities in software engineering for robotics. At the workshop, we held a panel on best practices for robotic software engineering, which featured experts of the NASA/Caltech Jet Propulsion Laboratory (JPL), Bosch, Waymo, and XITASO. We invited the panelists to share their best practices with the community below.

## 2. Architecting Small-Scale Flight Software at JPL

The Small-Scale Flight Software Group at JPL designs, develops, integrates, and tests embedded software that runs on autonomous spacecraft, landers, and surface exploration vehicles. By tradition, JPL refers to this software as "flight software" or FSW, even though not all of it actually flies (e.g., some of it drives on the surfaces of planets). The Small-Scale Flight Software Group focuses on smaller-scale missions such as CubeSats, SmallSats, and planetary rovers. Recent missions supported by the group include the ASTERIA space telescope [1] and the Ingenuity Mars helicopter [4]. Upcoming missions include Lunar Flashlight [5] (a mission to the moon) and Near-Earth Asteroid Scout [6] (a mission to a near-Earth asteroid). This section of the paper discusses the approach to architecting flight software used in the Small-Scale Flight Software Group at JPL. The discussion focuses on F Prime [3], an open-source multi-mission framework for flight and embedded software developed by the group.

### 2.1 Essential Quality Attributes for Mission Success

The Small-Scale Flight Software Group has identified the following FSW quality attributes as essential for mission success:
- **Analyzability:** The FSW should be easy to analyze, in order to detect any correctness or performance issues.
- **Efficiency:** The software should use memory and CPU efficiently and should fit within available resources.
- **Modularity:** The software should be clearly and logically decomposed into discrete units of function (modules).
- **Operability:** The software should be easy to use by the operators, i.e., the personnel who carry out the mission.
- **Testability:** The software should be easy to test at the module level and at the system level.
- **Understandability:** The software design should be easy to understand and communicate.

The group has also found that a high degree of software reuse is essential, especially for small-scale missions with constrained schedules and budgets. The group's engineers believe that the best way to achieve reuse is to use a strong multi-mission framework. When developing such a framework, the following quality attributes are important:
- **Adaptability:** The framework should be adaptable to new uses, beyond the specific uses known to the framework designers.

---

[2] https://rose-workshops.github.io

- **Configurability:** Key parameters of the framework should be changeable without modifying the framework code. For example, fixed data sizes should be configurable for different projects with different requirements and resources.
- **Portability:** The framework should be readily portable to new computing platforms.
- **Reusability:** The framework should provide core software components that are reusable without modification.
- **Usability:** The framework should be easy to understand and use.

## 2.2 F Prime

The goal of F Prime is to provide the quality attributes described in the previous section, thereby enabling the rapid development of high-quality flight software with comparatively low cost and high reuse [5]. F Prime provides the following elements: (1) an architecture that decomposes flight software into discrete **components** with structured communication based on **ports**; (2) a C++ framework providing core capabilities such as message queues and threads; (3) tools for specifying components and connections and automatically generating code; (4) a growing collection of ready-to-use components; and (5) tools for testing flight software at the unit and integration levels. F Prime meets the quality attributes stated above as follows:

1. **Analyzability:** Each F Prime application has a formal model that expresses a **topology**, i.e., a set of component instances and their connections. The Small-Scale Flight Software Group has developed a simple domain-specific modeling language called FPP [2] for expressing F Prime models. FPP checks the model for basic correctness properties like (a) whether ports are connected and (b) whether the direction, port numbers, and type of a connection are correct. In the future, FPP models can support more sophisticated analysis, e.g., performance analysis.
2. **Efficiency:** The F Prime framework is written in C++. The coding style is efficient: for example, it uses references instead of copying data where possible. The point-to-point communication through ports avoids the overhead associated with a software message bus.
3. **Modularity:** The decomposition of F Prime applications into components provides good modularity.
4. **Operability:** F Prime provides support for standard operations features such as commanding, telemetry, event reporting, and parameters (system constants that may be updated on command).
5. **Testability:** F Prime has a code generator that supports unit testing at the component level. F Prime also has a ground data system (GDS) for interactive testing of a complete FSW application (integration testing). Finally, F Prime has tools for writing automated integration tests.
6. **Understandability:** The F Prime modeling approach makes the design understandable. For example, it is easy to observe (a) the data carried on each port; (b) the ports defined in the interface of each component; and (c) how the ports are connected in the topology.
7. **Adaptability:** F Prime provides a set of common patterns (e.g., direct function calls, message passing, periodic behavior, asynchronous behavior). The patterns are flexible, and users can combine them in custom ways.

8. **Configurability:** Configuration occurs through special files that provide default values and that users can override.
9. **Portability:** F Prime has a serialization API providing a platform-independent way to store values as binary data. An OS abstraction layer hides the details of each platform and is portable to new platforms.
10. **Reusability:** The component-based approach supports reuse. Component interfaces are well-specified and conform to standards imposed by the framework. Components have no direct compile- or link-time dependencies on each other, so it is easy to arrange and rearrange the same components into different topologies.
11. **Usability:** FPP makes it easy to write F Prime models and to use them to generate code. Code generation enhances usability by minimizing the amount of repetitive code that users must write.

JPL has used F Prime on several flight projects, including the projects mentioned at the start of this section. These projects show that F Prime does in fact provide the quality attributes stated. The focus of F Prime to date has been on small-scale systems. However, the Small-Scale Flight Software Group believes that the architecture should scale to larger systems.

# 3. Model-Driven Engineering of Robotics Software at Bosch

Model-driven engineering (MDE) techniques still today have a tough stance with engineers and developers in several industrial domains, including robotics. This section presents some of the potential causes and hints at applications of MDE in production at Bosch that nevertheless are common or gaining acceptance.

## 3.1 Reasons Hindering MDE Adoption in Production

Quality attributes taken as criteria to decide for a particular MDE method or tool vary considerably depending on the state of design or development, i.e., the product development phase. Factors such as **Extensibility** (number of extensions points, **Modularity**, **Reusability**) that are of importance in early project phases, might be less important in later project stages.

In a potential transfer or hand-over of projects from research and innovation to production, **Understandability** (readability, writability) and **Explainability** of models as well as their **Adaptability to** specific domain problems and use-cases get increasingly important.

In later product phases, i.e., in production or operation, factors get relevant that were less relevant before. These include, among others, their potential to being scaled (to larger teams through available training material, to distributed teams, …), available expertise among developers, licensing and costs per developer, availability of professional and potentially qualified tools (editors, code generators, ...), support for release of safety-critical applications, and — last but not least — technical alignment with established tools and development processes of the organization.

In summary, while there is a lot of freedom of choice in early phases, organizational reasons dominate in later stages. As a consequence, MDE methods and tools find their way into industrial practice only if they either i) are powerful and flexible enough to span the mentioned development phases or ii) provide strong support and convincing arguments for a particular phase.

## 3.2 Levels of Adoption

Below we report the various levels of adoption of MDE technologies at Bosch.

**Established -** Established, yet not necessarily best practice, are MDE methods revolving around using general-purpose modeling languages (GPML) like UML and SysML for the sake of documentation and communication. Main reasons for their use are, among others, (1) broad availability of documentation and training material, (2) fair group of developers with at least limited experience, (3) availability of professional, qualified tools, and (4) use of GPML artifacts within design documentation and assurance cases [7]. However, since several others of the potential arguments for the use of MDE stated above are not true for this setup, these solutions often lead to great frustration with developers. Developers often don't gain significant advantage/support from these solutions for their daily work, but rather perceive them as a bothersome necessity.

**Common - Common** uses for MDE methods in robotics systems are reasonable abstractions for complex, domain-specific problems, such as behaviors and complex control. For behavior modeling, behavior trees and similar abstractions became popular for robotics applications, driven by their extensive tool support and wide adoption, e.g., in the gaming industry.

**Advancing -** Increasingly used are lightweight techniques that allow modeling right inside or next to code, e.g., annotations in code comments or integration with markdown that allow extraction of models from code repositories. These techniques often provide the advantage of (1) being easy to integrate into existing development environments (e.g., usable in text editor of choice), (2) being familiar to developers, and, foremost, (3) decreasing the danger of models and code diverging (code with comments and annotations being the single source of truth). Depending on the aspects modeled in code, design documentation can be generated (cf. plantUML), traceability with requirements can be established, and in the end required artifacts for product release and assurance cases can be generated.

Overall, the easier it is to integrate MDE into existing development environments and different development phases, and the more the persons that should provide the model are supported in their daily work, the more MDE has a chance to enter industrial practice.

# 4. Testing and Simulation of Autonomous Driving Software at Waymo

Waymo is developing and commercializing an SAE L4 [11] autonomous driving system (ADS) called the Waymo Driver. The Waymo Driver is an integrated system of enhanced sensing technologies including Lidar, Radar, and cameras, and onboard software. Through Waymo's safety methodologies [12] and testing, Waymo has deployed the world's first fully autonomous ride-hailing service (i.e., ADS-operated) service in Phoenix, Arizona called Waymo One. Due to the complex nature of interactions of an autonomous vehicle with the world, a variety of techniques are used for development and testing. The methods involve both physical and software testing in a variety of complementing ways. Closed course testing provides system-level testing of both basic and rare scenarios, public road testing provides additional discovery and validation, and simulation testing provides a safe way to test performance.

## 4.1 Testing at Waymo

Waymo's safety methodologies focus on the development, qualification, deployment, and sustained field operation of a Level 4 ADS with no human driver present [13]. Waymo utilizes several types of system-level testing to ensure the safety of the base vehicle and the safety of their autonomous operations. The methodologies used address the full life cycle of the autonomous vehicle, spanning from design and development through deployment and eventual decommission. System-level testing is conducted through three basic types: closed course, public road, and simulation.

**Closed course** testing is conducted on private test tracks. These tests are used to stage complex and rare scenarios in a safe and controlled environment. In total, Waymo has completed over 40,000 unique scenarios in closed course environments. These scenarios are used in their safety methodology to determine system performance against rare challenging events that would be hard to capture otherwise.

Additionally, the miles driven on **public roads** are extremely valuable for measuring the safety and performance of the ADS. The 20 million autonomous miles Waymo has driven in 25 cities across the US are used to test and validate the system. New and rare scenarios that are discovered can be added to closed-course testing, increasing coverage. The behavior observed in these miles is also used to validate the simulation. Miles driven are also used to calculate direct empirical testing of ADS competencies.

Finally, Waymo's **simulation** systems are used to test ADS capabilities at a large scale. Using simulation, Waymo engineers can test thousands of scenarios and road user interactions that would take far longer to experience in the real world. It also allows for testing scenarios that are difficult to reproduce in closed-course testing in a safe and reliable way. Simulation is also used as a source of regression testing, ensuring that ADS capabilities remain at high quality.

## 4.2 Simulation at Waymo

### 4.2.1 Simulated Scenarios

Simulation is used to recreate a variety of scenarios and is designed to test both basic behavior competencies and advanced functionalities. These scenarios are harvested from a variety of sources including driving logs, naturalistic driving research data, and reconstructed crashes [13].

The simulation platform used for these reconstructions is designed to provide a virtual testing environment that serves as a digital twin of the real-world driving environment. The simulation platform includes simulation of sensors, physics, and the behavioral layer to produce an accurate representation of reality.

First, sensors are simulated to provide realistic perception performance. While the scenario is described by the global positions of all relevant actors at all times, the ADS sensors and positioning limit its awareness. The sensor simulation accounts for sensor range, the field of view, sweeping behavior, and latency. This allows the simulation to accurately represent the time to perceive objects based on three-dimensional scene elements such as off-road obstructions. It also replicates the timing profile of the onboard system. Additionally, the simulated performance is validated in onboard tests to ensure realism. In this way, the simulation can faithfully reproduce the behavior of the onboard system.

Then, the Waymo Driver's behavior layer is executed within the simulation platform to control the simulated vehicle. The same behavior logic used in the on-road ADS deployment is tested in the simulation logic using the simulated sensor data, map, and vehicle dynamics to perform the driving task. The driving performance is then evaluated with the simulated data and judged for its safety and quality against a benchmark.

### 4.2.2 Large Scale Logs

Waymo maintains a large bank of historical logs collected during the development process. These logs allow engineers to discover issues in new situations, reveal previously unknown issues, and allow for the calculation of metrics on real driving data. Waymo automatically simulates millions of miles of driving data with new software release candidates. Additionally, engineers can simulate miles on demand and look at relevant events or changes caused by their code.

Log-based simulation can either be run in *closed-loop* or *open-loop*. The closed-loop simulation begins at a specific instance in time of the log and then continues using simulated data from that point. This kind of simulation tests how the software reacts to a scenario and is used to judge both the quality and safety of the ADS. Open-loop simulation simulates a single point in time in the log, without evolving the simulation between time steps. This type of simulation is used to test the software for reaction to a unique set of inputs and can be easily compared to a reference input. Simulating against a large set of logs leads to higher confidence in the resulting behavior. However, the size of the simulation precludes an individual inspection of each result. In order to deal with the scale, analyzers that can comb through the resulting data for interesting or applicable cases are used, along with metrics that measure aspects of driving performance such as adherence to road rules.

## 5. Testing Robotics Software at XITASO

Software testing is "the effort of executing programs with the intention of revealing failures" [8]. While maybe not covering all modern interpretations of testing, the definition is suitable to explain the XITASO approach on effective and efficient testing of robotics software.

At XITASO, engineers operate based on the belief that *software engineering is people business*. Software is made for people (that use it embedded in some machine) by people (with strong machine support) who decide what the software should do, how interactions should look like, and how it should be structured internally. Hence, *software testing is people business* as well. Central decisions must be made by humans, as they cannot be decided by a machine alone, for example:

- Selecting appropriate test cases,
- Setting up appropriate test environments,
- Analyzing test results, or even harder: defining the expected behavior of a program in a test case (i.e., the test oracle problem).

At the same time, XITASO engineers are convinced that in software development, they should *never send a human to do a machine's job*[3]. Hence, everything that can be efficiently automated should be done automatically by a machine. At first glance, this may seem like a contradiction to *software engineering is people business. However,* it is not: by delegating repetitive, clearly specified tasks to machines, people can preserve their mental capacity to make the important (and often necessarily creative) decisions in software engineering and testing. What does this mindset imply for the XITASO approach on testing robotics software?

Test case selection should be guided by a focus on the most valuable or risky parts of the system. The value of a system can be best judged together with actual users and stakeholders – recall the "people business". For this reason, acceptance testing [9] is very widely used throughout XITASO projects. In robotics projects, the acceptance tests usually involve complete hardware setups, including robots, periphery and an appropriate target environment, e.g., some part of a production site. Risky system parts may be determined from various considerations. In robotics, operational safety is a concern, as damage caused by the system is a huge risk. Safety-related analyses, like FMEA, can be a valuable source of test cases. Also valuable is experience derived from the development or operation history: a functionality in the system that is known to break easily, e.g., a force-based manipulation task, should be tested thoroughly and considered for regular regression testing [9].

Setting up appropriate test environments is a particularly delicate task in robotic software development, as many test cases depend heavily on the appropriate state of the physical environment. Based on the experience of XITASO engineers, a mixed approach to testing involving various environments and appropriate test activities is preferable: for automated integration tests during development, an environment that involves hardware mocking or basic simulations is a better fit than actual hardware tests. In large scale projects, more elaborate and detailed simulations can be appropriate. This removes hardware-related complexity from developers and testers. However, acceptance testing should be done in an environment as close to the productive scenario as possible. It can be worthwhile to have additional test environments that involve only parts of the relevant hardware: for bin-picking applications it makes sense to set up a specific test environment for the sensor that recognizes the objects to be picked. Such a dedicated test environment with limited hardware scope is often easier to set up automatically compared to a full acceptance testing environment on system level. At XITASO, the decision which mix of environments and test activities is most suitable in a given context is made by the development team together with other relevant stakeholders.

---

[3] as put by Agent Smith in the movie „The Matrix"

Judging whether the result of a test case execution is success or failure is also often hard for a machine, especially on the upper levels of the test pyramid (integration or end-to-end level). During development of a 3D visualization of safety zones for an industrial robot controller at XITASO, end-to-end testing was done manually, partly because correctness of a 3D visualization is relatively easy to verify for humans. The most efficient work split in such cases is to leave judgement of results to humans, but to automate as much other testing work as possible - e.g., installing different versions of the robot controller software, configuring system parameters and archiving logs generated during the tests. Thanks to the DevOps movement [10], those activities can be efficiently automated. There exist various virtualization/containerization approaches to ease automatic installation of software components, applicable also to real-time critical systems like robot controllers. Furthermore, there are methods and tools for treating configuration and infrastructure setup as code. These techniques, when applied to testing activities, allow for (i) easy automation of tedious tasks and (ii) the creation of clearly defined and traceable test environments and test case protocols. Moreover, modern CI/CD environments provide the flexibility to integrate machine and human activities in a defined workflow. This way, machines can do the tedious, repetitive parts of testing and leave other, more complex tasks – e.g., inspecting 3D safety zone visualizations – to human testers.

The DevOps movement largely builds on automation of repetitive tasks – and this is the key to solve the tension between *software testing as people business* and *never sending a human to do a machine's job*. At XITASO, engineers use today's computing power, tools and methods to create testing and delivery pipelines that are precisely defined, traceable, and repeatable. This way, they allow humans to do what machines cannot do: to judge the quality of complex software-intensive systems like robots.

## 6. Conclusions

In this article we elicited the software engineering practices applied by JPL, Bosch, Waymo, and XITASO when working on their robotic systems. They provided insights and a concrete overview about the state of the practice of robotics software engineering, touching upon the architecture of the system, testing, simulation, and maintenance. We hope that the practices elicited in this paper will help professionals in building on these lessons learned.